\documentclass[12pt,notoc,final,nohyper]{JHEP3}
\usepackage{amsfonts}
\usepackage{amssymb}
\usepackage{graphics}
\def\a{\alpha}

\def\ad{{\dot{a}}}
\def\ab{{\alpha\beta}}
\def\cd{{\gamma\delta}}
\def\b{\beta}

\def\bm{\mathbb{M}}
\def\cp{\mathbb{CP}}
\def\cn{{\cal N}}
\def\d{\delta}

\def\eps{\epsilon}
\def\g{\gamma}
\def\l{\lambda}

\def\s{\sigma}

\def\th{\theta}

\def\Th{\Theta}

\newcommand{\be}{\begin{equation}}
\newcommand{\ee}{\end{equation}}
\newcommand{\bea}{\begin{eqnarray}}
\newcommand{\eea}{\end{eqnarray}}
\newcommand{\nn}{\nonumber\\}

\preprint{ITFA-2004-42\\hep-th/0410014\\}

\title{A Six Dimensional View on Twistors}

\author{Annamaria Sinkovics and Erik Verlinde\\
Institute for Theoretical Physics, University of Amsterdam \\
Valckenierstraat 65, 1018 XE Amsterdam, The Netherlands}

\abstract{We study the twistor formulation of the classical ${\cal N}=4$
super Yang-Mills theory on the quadric submanifold of 
$\cp^{3|3} \times \cp^{3|3}$. We reformulate the twistor equations in six 
dimension, where the superconformal symmetry is manifest, and find a connection
to complexified $AdS_5$.}

\begin{document}

\section{Introduction}
Recently there has been a renewed interest in twistor formulations of gauge 
theories.
By studying the structure of maximally helicity violating amplitudes
Witten has constructed a new formulation of ${\cal N} =4$ gauge theory 
\cite{wittenmhv}.
The structure of these amplitudes becomes apparent when transforming from 
momentum to twistor space,
where the amplitudes are supported on holomorphic curves. Follow-up works 
extended this result for
non-maximally helicity violating amplitudes \cite{f1},
for ``googly amplitudes'' (i.e. amplitudes in the opposite helicity 
description)
\cite{f2}, for analyzing loop amplitudes \cite{f12} and gravity amplitudes 
\cite{f8}.
For further extension of the formalism and for advances on the diagrammatics 
of amplitude computations see \cite{f3}.

These results can be interpreted \cite{wittenmhv} by formulating ${\cal N} 
=4$ gauge theory as a topological string
theory, the B-model with target space of the supermanifold $\cp^{3|4}$. In 
this way the structure of perturbative
Yang-Mills amplitudes arises by including the contribution of D-instantons. 
Alternative string formulations for
describing the perturbative ${\cal N}=4$ twistor space amplitudes has also 
been proposed in \cite{alt}, see also \cite{siegelc}.
In this note we are interested in a different formulation of the ${\cal 
N}=4$ SYM, proposed by Witten \cite{wittenold}.
According to \cite{wittenold}, the full classical
Yang-Mills field equations, not just the self-dual or anti-self-dual part, 
can be constructed in terms of a vector bundle on
a quadric submanifold $Q \in \cp^{3|3}\times \cp^{3|3}$.
This formulation generalizes Ward's construction \cite{ward} of 
(anti)self-dual gauge fields from
vector bundles on a single $\cp^3$. A concise summary of Ward's 
formulation is given in the appendix of
\cite{wittenmhv}. Unlike the formulation on $\cp^{3|4}$, the construction 
from $Q\in \cp^{3|3}\times \cp^{3|3}$
is parity symmetric, a helicity flip exchanges the two $\cp^{3|3}$s.

The connection between Yang-Mills theory and the quadric $Q$ has gained 
further interest through the
recent duality conjectures in topological string theory. First, in  
\cite{vafa1} it was argued that
an S-duality relates the B-model on $\cp^{3|4}$ to the A-model on the same 
supermanifold, see also \cite{vafa2}.
Secondly, it was conjectured that by mirror symmetry the A-model on 
$\cp^{3|4}$ becomes the B-model on the
quadric $Q$ in $\cp^{3|3} \times \cp^{3|3}$.  A proof of this mirror map was 
presented in \cite{agvaf}.
The D-instanton contributions of the original B-model are mapped first 
through the S-duality
to perturbative A-model amplitudes. After the additional mirror symmetry we 
arrive again at the B-model, but
without the D-instanton contributions. This means the Yang-Mills equations 
could be directly
related to the B-model on the quadric, and may thus be formulated in terms 
of a holomorphic Chern-Simons theory on
this space. This possibility was already mentioned in \cite{wittenmhv}, but 
for a concrete realization of
this idea one needs to overcome the obstacle of finding a proper measure on 
the quadric.

It is interesting to note that, while the B-model on $\cp^{3|4}$ is related 
to weakly coupled ${\cal N}=4$ SYM,
the B-model on the quadric should again be give a strong coupled 
formulation, since it follows from the conjectured
$S$-duality for topological strings and mirror symmetry \cite{vafa1,vafa2}. 
Furthermore, the target space $Q_5$ of the 
B-model is a complex five-dimensional
supermanifold and is symmetric under the superconformal group $SU(4|3)$. It 
is therefore natural to ask whether there is a
connection with the $AdS_5\times S^5$, which is also dual to the same theory 
in the same regime. In this paper we will make a step in this direction by
reformulating
the (super-)twistor equations in a 6 dimensional notation that 
makes the superconformal invariance manifest.
In our formulation 4d Minkowski space will be identified with the lightcone 
embedded in the six dimensional flat
space modded out by rescalings. This projective version of 6d space has also 
other components, one of which can be identified
with $AdS_5$. The twistor equations rewritten in the 6d notation take a 
particularly simple form.

\section{Twistor construction of the Yang-Mills equations}
Let us first fix conventions.
We work in signature $\eta^{\mu \nu} = \{-, +, +, +\}$, and use
complexified Minkowski space $\mathbb{M}_4$ so that the coordinate $x \in 
\mathbb{M}_4$ are complex. In this paper we will write most equations explicitly 
in coordinates that are defined for non-compact Minkowski space, but our results can be extended to
its compactified version, which we also denote by $\mathbb{M}_4$. 
Undotted and dotted indices denote spinors transforming in the (1/2, 0) and 
(0, 1/2)
representations, and can be raised and lowered with the two-index 
antisymmetric tensor $\eps$.
In spinor notation
$$
x_{a \ad} = \s^{\mu}_{a \ad} x_{\mu} = x_0 \d_{a \ad} + \vec{x} \vec{\s}_{a 
\ad}.
$$
The bosonic twistor equation is written as \cite{wittenmhv}
\be
\label{twistoreq} V_{\ad} + x_{a \ad} V^{a} = 0, \quad\qquad \ad=1,2.
\ee
It can be viewed in two ways: given $x$,
it determines a curve in the space $\cp^3$, which is parametrized by the 
homogeneous coordinates $\l$ and $\mu$.
The space $\cp^3$ is called twistor space. The curve itself is a copy of 
$\cp^1$,
since the equation can be solved for $V^{a}$  of $V_{\ad}$, or the reverse. 
In the analysis of the scattering amplitudes,
this curve arises after Fourier transforming the amplitudes from momentum to 
twistor space.
After the transformation, the amplitudes are localized on the curve given by 
the twistor equation.
{}From another point of view, given $V^{a}$ and $V_{\ad}$, the twistor 
equation determines a two dimensional subspace in $\mathbb{M}_4$, called alpha-plane.
The twistor equation is naturally connected to the (anti)self-dual 
Yang-Mills equation via Ward's construction.
The basic idea of this construction is that the information of 
(anti)self-dual gauge fields can be encoded in the structure of complex 
vector bundles.
Consider a complex vector bundle over $\mathbb{M}_4$ with a 
connection on it.
In general, parallel transport with this connection is not integrable.
However, according to Ward's construction,
we have integrability when restricting to the alpha-planes, if and only if 
the gauge field is anti-self-dual.
The set of all alpha-planes is the twistor space $\cp^3$.
There is of course analogous construction for a self-dual gauge field, where 
the complex 2-planes of integrability are now
called beta-planes. Thus by
Ward's theorem, an (anti)self-dual gauge field corresponds to a  
vector bundle on the twistor space $\cp^3$, and
this vector bundle is trivial when restricted to a 2-dimensional subspace 
defined by the twistor equation.
It is natural to try to extend this construction to the full Yang-Mills
equations. While the self-dual gauge field equations are algebraic, the full 
Yang-Mills equations are a differential
equation. In \cite{wittenold} Witten achieves the extension by embedding 
Minkowski space in a bigger space,
$\mathbb{M}_4 \times \mathbb{M}_4$.  
By writing $x\in \bm_4$ as $x={1\over
2}(y+z)$  with
$$
(y,z)\in \bm_4\times \bm_4,
$$
one can split the 4d connection $\nabla_x$ as
\be
\nabla_x=\nabla_y+\nabla_z.
\ee
The original Minkowski space thus corresponds to the diagonal $y=z$ inside $\bm_4\times\bm_4$.  
The connection $\nabla_x$ satisfies the Yang-Mills field equations if 
$\nabla_y$ is anti-self-dual, $\nabla_z$ is self dual and both connections
 mutually commute. Thus we have
\bea
\left\lbrack \nabla_y,\nabla_y\right\rbrack +  \left\lbrack
\nabla_y,\nabla_y\right\rbrack^* &=&0\nn
\left\lbrack \nabla_z,\nabla_z\right\rbrack -  \left\lbrack
\nabla_z,\nabla_z\right\rbrack^* &=&0
\eea
and
$$
\left\lbrack \nabla_y,\nabla_z\right\rbrack =0.
$$
 From these relations it follows that
$$
\left\lbrack \nabla_x,\nabla_x\right\rbrack =  \left\lbrack
\nabla_y,\nabla_y\right\rbrack+\left\lbrack \nabla_z,\nabla_z\right\rbrack
$$
and finally the Jacobi identity implies
\be
\left \lbrack \nabla_x, \left\lbrack
\nabla_x,\nabla_x\right\rbrack^*\right\rbrack =
\left \lbrack \nabla_y+\nabla_z,
- \left\lbrack \nabla_y,\nabla_y\right\rbrack+\left\lbrack
\nabla_z,\nabla_z\right\rbrack 
\right \rbrack=0.
\ee
One of the main points of \cite{wittenold} is that a gauge connection can 
only be split in this way if it corresponds to a vector bundle on $\cp^3\times \cp^3$, 
again trivial on each $\cp^1\times \cp^1$. This is a very strong requirement that is
not  satisfied by a general solution of the Yang-Mills field equations. 
Every gauge field on $\bm_4$, not necessary satisfying any equation,
is equivalent to a  vector bundle on the manifold
\be 
\label{Qbos}
Q_5 = \left\{ (U, V) \in \cp^3 \times \cp^3 | 
\sum_{\alpha=1}^4 U_{\alpha}  V^{\alpha}=0 \right\}.
\ee
Here $\alpha=(a,\ad)$ is a four 
component spinor index. The space $Q_5$ has complex dimension 5 and can be viewed as the space of all 
lightlike lines in $\mathbb{M}_4$. The light like lines through a given point in $\bm_4$ form a $\cp^1\times\cp^1$ inside $Q_5$, with one $\cp^1$ in each factor of $\cp^3\times \cp^3$.  A vector bundle association with a gauge field 
on $\bm_4$ is trivial on every such $\cp^1\times \cp^1$. Gauge fields that satisfy the Yang-Mills equation $ D^* F =0$
corresponds to a vector bundle on $Q_5$ that can be extended to a small local neighborhood of $Q_5$
inside $\cp^3\times\cp^3$. To be precise, it is necessary and sufficient to extend the vector bundle from
$Q_5$ up to third order in a local Taylor expansion. This means that the vector bundle 
actually lives on a quadric given by $(U_\a V^\a)^4=0$, which can then be taken as the actual defining equation of $Q_5$. 
The extension of the connection to third order away from $Q_5$ also implies that the Yang-Mills gauge field can be extended to $\bm_4\times\bm_4$ away from the diagonal up to third order in the $w=y-z$. Ward's construction relates
a connection on $Q_5$ to an anti-selfdual connection $\nabla_y$ and self-dual connection $\nabla_z$, but to get the Yang-Mills equations these connections also have to commute in the neighbourhood of the diagonal. This is what leads to 
the above mentioned requirements.

\begin{figure}
\begin{center}
\includegraphics{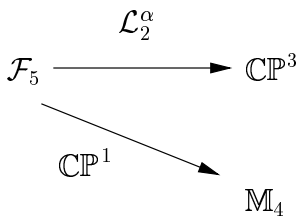}
\end{center}
\caption{Twistor space $\cp^3$ and Minkowski space $\mathbb{M}_4$ can be regarded as the base of a fiber bundle with total space ${\cal F}_5$. 
The corresponding fibers are the alpha-plane ${\cal L}^\alpha_2$ and $\cp^1$.}
\end{figure}

It is useful to compare this twistor construction of the Yang-Mills equations with the usual one for the 
(anti-)self dual equation in terms of a schematic diagram, as indicated in fig.1. 
Complexified Minkowski $\mathbb{M}_4$ and the usual twistor space $\cp^3$ can both be seen as projections of the same 
five dimensional space denoted by ${\cal F}_5$.
The $\cp^1$ fiber over $\mathbb{M}_4$  corresponds to the set of all alpha-planes through a given point. 
The two-dimensional fiber ${\cal L}^\alpha_2$ over $\cp^3$ is just the alpha-plane itself.  
Similarly we can construct a fiber bundle over $Q_5$ by taking 
the lightray ${\cal L}_1\sim \cp^1$ parametrized by a point in $Q_5$ as the 
fiber. The resulting total space ${\cal F}_6$ is also a fiber bundle over 
$\mathbb{M}_4$ with fiber equal to $\cp^1\times\cp^1$, which is the space of 
all lightrays through a given point on $\mathbb{M}_4$. This is shown in fig. 2.

\begin{figure}
\begin{center}
\includegraphics{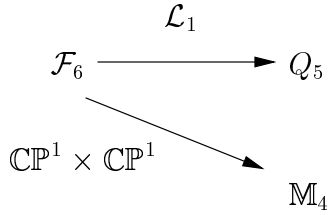}
\end{center}
\caption
{The total space ${\cal F}_6$ is fibered over the space of lightrays $Q_5$ with the lightray ${\cal L}_1$ itself as fiber.
Points in Minkowski space $\mathbb{M}_4$ lift to $\cp^1\times\cp^1$ fibers in ${\cal F}_6$. }
\end{figure}

\subsection{Supersymmetric extension}
As in the bosonic space, we can examine if the supersymmetric Yang-Mills 
equations are equivalent to an
integrability conditions along lightlike lines through a given point in the 
superspace.
It turns out that this is indeed the case for $\cn =3$ or $\cn=4$ 
supersymmetry
(these two theories are basically equivalent).
The supertwistor equations for the alpha-plane are  \cite{wittenmhv}
\bea
V_{\ad} + x^{R}_{a \ad} V^{a} &=& 0 \quad \ad=1,2 \nonumber
\\
\psi_{I} + \th_{aI} V^a &=& 0 \quad I = 1 \ldots {\cal N}. \label{supertwistor}
\eea
Here we introduced  anticommuting coordinates $\th_{a I}$, $I=1 
\ldots {\cal N}$ for
${\cal N}$ supersymmetries. The supertwistor space $\cp^{3|3}$ is thus
parametrized by
$(V^a, V_{\ad}, \psi^I)$, where the $\psi^{I}$ are spinless anticommuting 
variables. Analogously, the beta-plane equations are given as
\bea
U_a +  x^L_{a\ad } U^{\ad} &=& 0  \nonumber \\
\eta^{I}+\th_{\ad}^I U^{\ad} \label{st2}  &=& 0.
\eea
The pair of equations (\ref{supertwistor}) for the alpha-planes can, as in the bosonic case, be found by a 
partial Fourier transformation of the MHV amplitudes to twistor space. 
Similarly, the beta plane equations (\ref{st2}) arise by
a Fourier transformation of the MHV amplitudes with opposite helicity.
It turns out, however, that the $x$-coordinate that appears in these equations is different for the left-handed
and right-handed helicities, hence the superscript.

The alpha- and beta-planes are invariant under supersymmetry and 
superconformal variations. The supersymmetry variations which leave the set of alpha-plane equations 
invariant are
\bea
\label{susyR}
\delta x^R_{a\ad} &=& - \eps^I_{\ad}\theta_{aI}\qquad \delta 
\theta_{aI}=\eps_{Ia} \nn
\delta V_{\ad} &=&- \eps^I_{\ad} \psi_I\qquad \delta \psi_I =-\eps_{a I} 
V^a. \nonumber
\eea
The alpha-planes are also invariant under the superconformal variations
\bea
\label{Rvar}
\delta x^R_{a\ad} &=& x^R_{b\ad}\widetilde{\eps}^{bI}\theta_{aI},\qquad\quad
\delta \theta_{aI}=\widetilde{\eps}_{I}^{\ad} x^R_{a\ad}-
\widetilde{\eps}^{bJ} \theta_{b J} \theta_{aI} \nn
\delta V^a &=& \widetilde{\eps}^{aI}\psi_I,\qquad\qquad \delta \psi_I
=\widetilde{\eps}_{\ad I} V^{\ad}. \nonumber
\eea
The supersymmetry and superconformal variations which leave the beta-planes 
invariant
are completely analogous.
By comparing the transformation rules we find the chiral and anti-chiral 
coordinates are related as
$$
x^R_{a \ad} = -x^L_{a\ad} + \th_{aI} \th_{\ad}^{I}.
$$
Super lightrays are obtained by intersecting the alpha- and beta-planes. Imposing both the 
alpha-plane equations (\ref{supertwistor}) as well as those for the beta planes (\ref{st2}) leads to a condition on the
supertwistors. Namely, one only obtains a non-trivial solution provided that 
\be
U_{a} V^{a} +  U_{\ad} V^{\ad} + \psi_I \eta^{I} =0.
\ee
This defines the generalization of the manifold $Q_5$ to the supersymmetric situation.
\be 
\label{Q}
Q_{5|2N} = \left\{ (U,\eta, V,\psi) \in \cp^{3|N} \times \cp^{3|N} | 
U_{\alpha}  V^{\alpha}+\psi_I\eta^I =0 \right\}.
\ee
The quadric submanifold $Q_{5|2N}$ is the space of all supersymmetric lightlike lines \cite{ferber}. 
Just as in the bosonic case one can define a fibre bundle over it
with total space ${\cal F}_{6|4N}$, which projects down on $\bm_{4|4N}$ along $\cp^1\times\cp^1$ fibers, see fig. 3.

The supersymmetric lightlike lines ${\cal L}_{1|2N}$, unlike in the bosonic case, are not
one dimensional, thus integrability along them is not any more a trivial
condition. According to \cite{wittenold},
for  $\cn=3$ supersymmetry the integrability on the quadric corresponds to 
the $\cn =3$ supersymmetric equations
of motion.  For $\cn=4$ supersymmetry an additional condition is necessary,
see \cite{wittenold}.  
Henceforth, in the rest paper we will take ${\cal N} =3$. 
Ward's construction relates a vector bundle on $Q_{5|6}$
to a supergauge field on supersymmetric Minkowski space $\mathbb{M}_{4|12}$.
It is an interesting open problem whether $N=3$ SYM can be reformulated 
as the holomorphic Chern-Simons theory corresponding to a B-model topological 
string on $Q_{5|6}$.
The quadric  $Q_{5|6}$ is a Calabi-Yau supermanifold, thus the B-model in rinciple can 
be constructed on it, and as mentioned before, the existence of the B-model on the 
quadric is also supported by a conjectured duality chain. The main problem in formulating the holomorphic Chern-Simons theory appears to be the definition of the appropriate measure.
This question will not be addressed in this paper. Our main concern is the reformulation of the twistor 
equations in a manifestly superconformal fashion, and in this way clarify the role of the quadric in super-Yang-Mills.

\begin{figure}
\begin{center}
\includegraphics{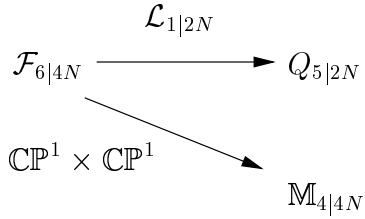}
\end{center}
\caption
{The total space ${\cal F}_{6,4N}$ is fibered over the space of supersymmetric 
lightrays $Q_{5|2N}$ as well as over 
supersymmetric Minkowski $\mathbb{M}_{4|4N}$. The corresponding fibers are 
the superlightray ${\cal L}_{1|N}$ and $\cp^1\times\cp^1$.}
\end{figure}

\section{The twistor equations in six dimensional notation}

The 4d Minkowski space $\bm$ can be identified with the set of all lightlike 
directions in a 6d flat space-time
${\mathbb M}_{4,2}$ with signature $--++++$ and metric
\be
ds^2=-dX^+dX^-+ dX_\mu dX^{\mu}, \qquad\quad \mu=1 \ldots 4.
\ee
Specifically, a lightlike direction in $\mathbb{M}_{4,2}$ can be 
parametrized as
\be
(X^+,X^-,X_{\mu})\sim (1,x^2, x_{\mu}),
\ee
where $x_{\mu}$ are the coordinates for a point in
$\bm$ and $x^2$ its length.The 6d isometry group $SO(4,2)$ acts as the 
conformal group on $\bm$.
Twistors are naturally formulated in this six dimensional language. In this 
way, a twistor corresponds
to a chiral spinor
\be
V^{\a} = \left(\begin{array}{c}  V^{\ad} \\  V^{a} \end{array} \right)
\ee
transforming in the $\mathbf 4$ of $SU(2,2)$. For every point in the 6d 
space-time we can define the antisymmetric matrix
\be
X_{\alpha\beta}\equiv
\left(\begin{array}{cc}X^+ \epsilon_{\dot{a}\dot{b}} &  X_{b\dot{a}} \\ 
-X_{a\dot{b}}& -X^-\epsilon_{ab}\end{array}\right)
\ee
where we used the spinor notation introduced before, so $X_{a {\dot b}} 
=\s^{\mu}_{a {\dot b}} X_{\mu}$.
The twistor equation then takes the simple 
form
\be
\label{a1}
X_{\alpha\beta}V^\beta=0
\ee
The first component of this equation
gives
\be
X^+ V_{\dot{a}}+X_{a\dot{a}} V^a=0.
\ee
By using the scale invariance to define $x_{a \ad}  = X_{a \ad} / X_{+}$,
one recognizes the twistor equation (\ref{twistoreq}).
The second component gives an additional equation, which is equivalent to 
the twistor equation provided
$$
-X^+X^- + X_{\mu} X^{\mu}=0.
$$
This is the lightcone condition on the six dimensional embedding space.
We thus recover the standard twistor equation describing the 
anti-self-dual alpha-plane in $\bm$.
In the six dimensional space $\bm_{4,2}$ where rescalings are not modded 
out, the twistor equation
defines an anti-self-dual null 3-plane through the origin. We will also call 
this alpha-plane.
Similarly there are beta-planes that are self-dual and that can be 
described via a similar
twistor equation but with the dotted and undotted indices interchanged. 
Specifically, we can raise
indices as
\be
\label{XD}
X^{\alpha\beta}={1\over 2}\epsilon^{\alpha\beta\gamma\delta}X_{\gamma\delta}
\ee
where $\epsilon^{a\dot{b}\dot{a} b}= \epsilon^{\dot{a}\dot{b}}\epsilon^{ab}$ 
(other entries follow by permutation).
The twistor equation for beta-planes then takes the 
form
\be
\label{b1}
X^{\alpha\beta}U_\beta=0
\ee
Imposing both type of twistor equations amounts to intersecting the 
alpha- and beta-planes. Generically these only intersect in the origin of the 
6d space. In Minkowski space this means there is no intersection at all. To 
get a nonzero intersection one should impose that
\be
U_\alpha V^\alpha=0.
\ee
This can be seen most easily in components:
$X^+ U_a V^a=-U^{\dot{b}}X_{a\dot{b}}V^a=-X^+U^{\dot{b}}V_{\dot{b}}$.
The intersection of an alpha-
and beta-plane yields in this case a null two-plane through the origin in 
6d, and corresponds to a lightray in 4d.

\subsection{Superconformal invariance}

We now proceed with the supersymmetric extension of the alpha- and 
beta-planes on the quadric.
The starting point is that the set of supertwistor equations has to be 
invariant under the superconformal group
in six dimension, $SU(4|3)$.  The superconformal group acts linearly on 
$(U_{\a},\eta_{I})$ and on $(V^{\a},\psi^{I})$,
which transform in the conjugate representations $(4|3)$ and 
$(\overline{4}|\overline{3})$.
Since the R-symmetry group is $U(3)$, the  $\eta_{I}$ and $\psi^{I}$ are in 
different representations.
Taking the product of the $(4|3)$ with itself we look for the anti-symmetric 
combinations.  The antisymmetric $6$
is identified with the $X_{\alpha\beta}$,  while the antisymmetric odd $12$ 
are the $\Theta_\alpha{}^I$.
The representation is completed by taking the symmetric combination $6$ of 
$(3\times 3)$, let us call this $\Phi_{IJ}$.
The symmetric field $\Phi_{IJ}$ can be thought of as a metric, which can be 
used to raise and lower the $U(3)$ indices.

The superconformal generators satisfy
$$
\lbrace Q_{\alpha}^{I}, Q^{\beta}_{J} \rbrace=M_{\alpha}{}^{\beta} 
\d^{I}_{J} + \d_{\a}{}^{\b} M^{I}{}_{J}
$$
where $M_{\a}{}^{\b}$ are the Lorentz generators, here written as a matrix in the fundamental 
representation of $SU(4)$, and the $M_I{}^J$ the $SU(3)$ 
R-symmetry generators.
The superconformal generators contain both the supersymmetry and additional 
conformal generators, which in the
four dimensional notation appeared separately. For the invariance of the 
twistor equations under the superconformal
group, we put the equations in the smallest possible representation $(4|3)$ 
of the superconformal group $SU(4|3)$.
We obtain the four even and three odd equations
\bea
X_{\a \b} V^{\b}  + \Theta_{\a}{}^J \psi_J &=& 0  \label{te} \\
\Th_{\a}{}^I V^{\a} + \Phi^{IJ} \psi_{J} &=& 0  \label{to}.
\eea
This set of equations describes the super alpha-planes, and are invariant 
under the action of the
superconformal generators. Indeed, examining the action of the generators on 
the set of twistor equations,
we find that (\ref{te}) is annihilated by $Q_{\a}^{I}$, and is mapped into 
(\ref{to}) under $Q^{\b}_{J}$.
The odd equations (\ref{to}) are mapped to the even equations (\ref{te}) 
under $Q_{\a}^{I}$, and are annihilated by the
$Q^{\b}_{J}$.
The beta-plane equations can be derived by analogous reasoning and
read
\bea
\label{6dbeta}
U_{\a} X^{\a \b} + \eta^{J} \Theta^{\b}{}_{ J}  &=& 0 \nn
U_{\a} \Theta^{\a}{}_{ I}  +\eta^{J} \Phi_{JI}  &=& 0.\label{bb}
\eea
The relation between the alpha- and beta-plane equations  will be further clarified in the following section.

We end this section by explaining how to get the supertwistor equations in 
the 4d notation of section 2. We only consider the equations for the alpha-planes. 
To reduce (\ref{te}) and (\ref{to}) to the 4d twistor equations (\ref{supertwistor}) one first converts
the four component indices $\alpha$ and $\beta$ to two-component notation. 
In addition, one has to identify the 4d coordinates
$x^R_{a\ad}$ and $\theta_\ad $ in terms of the 6d ones.
This turns out to be trickier than one might have expected at first: $x^R_{a\ad}$ and $\th_\ad$ 
are non-linearly defined in terms of the 6d coordinates. 
With hindsight this is not surprising because the 4d superconformal symmetries act non-linearly on the 
coordinates, while the action on the 6d coordinates is linear. After a bit of straightforward but tedious algebra one finds that the correct identifications are
\bea
x^R_{a\ad} &= & \left(X_{a\ad}-\Th_a{}^I \Th_{\ad I}\right)/ (X_+-{1\over 2}\Th_\ad{}^I\Th^\ad{}_I)\nn
\th_{\ad}^I & = & \Theta_\ad{}^I-x^R_{a\ad}\Theta^{aI}
\eea  
The coordinate $\Phi^{IJ}$ was used to raise and lower indices, and disappears in the 4d picture.
The reduction of the beta-plane equations (\ref{6dbeta}) to four dimensional notation proceeds analogously. 

\section{Combining alpha- and beta-planes: connection with $AdS_5$}

The twistor equations for the alpha- and beta-planes are scale invariant 
in 6 dimensions. One can add a point at infinity by introducing another coordinate chart obtained by the inversion map
\be
X_{\alpha\beta}\to -4\zeta^2 {X_{\alpha\beta}\over X^2}\qquad\qquad 
X^2={1\over 2}\epsilon^{\a\b\g\d}X_\ab X_\cd
\ee
where $\zeta$ is an arbitrary scale.
Inversion sends the light cone $X^2=0$ to the point at infinity. 
We will argue that it also exchanges the alpha- and beta-planes.
We first consider the bosonic twistor equations.  We use the parameter $\zeta$ to
modify the bosonic twistor equation as follows
\be
\label{xx}
X_\ab V^\b=\zeta U_\a.
\ee
Further, instead of the light cone we consider the five dimensional submanifold
\be
\label{AdS}
X^2=-4\zeta^2
\ee
This equation describes a complexified (Anti-)de Sitter space.
What we have achieved by introducing the parameter $\zeta$ is that on the 
submanifold (\ref{AdS}) the modified alpha-plane equation (\ref{xx}) is 
equivalent to the analogous equation for the beta plane 
\be
X^{\ab}U_\b=\zeta V_\a
\ee
This follows from the identity 
$$
X^{\ab}X_{\b\g}= \zeta^2 \delta^\a{}_\g
$$
where we made use of (\ref{AdS}). When $\zeta\neq 0$ we have the freedom 
to rescale $X$ and put $\zeta=1$.  In this case the submanifold (\ref{AdS}) is
the fixed locus of the inversion map. Furthermore, in the limit $\zeta\to 0$ 
one recovers the `old' equations  (\ref{a1}) and (\ref{b1}), and the 
AdS-manifold reduces again to the light cone. The above procedure is 
analogous to the way in which the massive Dirac equation reduces to two decoupled Weyl equation for the left- and right-handed components of a spinor. In this analogy the parameter $\zeta$ is as a ``mass'', and
$V$ and $U$ the left-handed and right-handed Weyl spinors.

This construction can be generalized to the supertwistor equations.
The modified form of the supersymmetric alpha-plane equations read
\bea
\label{XVU}
X_{\a \b} V^{\b}  + \Theta^{I}{}_{\a} \psi_I &=&  \zeta U_{\a}  \nn
\Th^{I}{}_{\a} V^{\a} + \Phi^{IJ} \psi_{J} &=&  \zeta \eta^{I}
\eea
where we added the supertwistors for the beta-plane equations on the right 
hand side, multiplied with an auxiliary parameter $\zeta$.
Note that the modification with $\zeta \neq 0$ is consistent with the $SU(4|3)$ superconformal 
symmetries. All of the variables appearing here are projective coordinates: 
one has the freedom to rescale $(U,\eta)$ and $(V,\psi)$ by arbitrary and independent complex 
variables. We can also rescale the coordinates $(X,\Th,\Phi)$ simultaneously with $\zeta$.
In principle this allows us to put the parameter $\zeta$ to an arbitrary value. 

The equations (\ref{XVU}) directly imply the quadric relation
\be
U_{\a} V^{\a} + \psi_{I} \eta^{I} =0.
\ee
as can be seen by replacing $U_\b$ and $\eta^I$ by the expressions on the 
l.h.s. We can modify the supersymmetric beta plane equations in a similar way 
\bea
\label{UXV}
U_\a X^{\ab}+\Th^\a{}_I\eta^I & = & \zeta V^\a, \nonumber \\
U_\a\Th^\a{}_I+\Phi_{IJ}\eta^J &= &  \zeta \psi_I.
\eea
We now require that these equations are consistent with (\ref{XVU}). 
This leads to a number of relations between the $X$, $\Th$ and $\Phi$ 
coordinates with upper and lower indices. Let us first take $\zeta\neq 0$. 
Note that both the alpha- and beta-plane equations can be 
written in matrix form by combining the coordinates as follows
\be
\label{matrix}
\zeta^{-1}\left(\begin{array}{cc} X & \Theta \\ \Theta & \Phi \end{array}\right).
\ee
The only difference between the alpha- and beta-planes is that in 
one case the indices are up and in the other case down, and more 
importantly that $(U,\eta)$ and $(V,\psi)$ are interchanged. 
It is now easy to see that the two sets of equations are consistent 
for $\zeta\neq 0$ if and only if the matrix (\ref{matrix}) is invertible, and its inverse is simply obtained by replacing upper with lower 
indices.  This leads to the relations
\bea
X^{\alpha\gamma}X_{\gamma\beta}+\Theta^\alpha{}_I\Theta_\beta{}^I&=& \zeta^2 \delta^\alpha{}_\beta \nn
\Theta^\alpha{}_I \Theta_\alpha{}^J+\Phi_{IK}\Phi^{KJ} &=& \zeta^2 \delta_I{}^J \nonumber
\eea
and
\bea
X^{\alpha\beta} \Theta_\beta{}^I+\Theta^\alpha{}_J\Phi^{JI} &=& 0 \nn
\Theta^\beta{}_I X_{\beta\alpha} +\Phi_{IJ}\Theta_\alpha{}^J&=&0. \nonumber
\eea
With these relations (\ref{XVU}) is equivalent to (\ref{UXV}), and hence it suffices to keep only one or the other set of equations.

To show the equivalence we assumed that $\zeta\neq 0$. But now we can take the limit $\zeta\to 0$ and obtain both the alpha- 
and the beta-plane equations.
In this limit
\be
X^{\a \b} X_{\a \b}-\Th^{\a}{}_I \Th_\a{}^I= 0
\ee 
This describes the super-lightcone in 6d. But when $\zeta\neq 0$ we find
\be
X^{\a \b} X_{\a \b}-\Th^{\a}{}_I \Th_\a{}^I=-4\zeta^2 
\ee 
Here we recognize a complexified supersymmetric version of $AdS_5$. 
The appearance of $AdS_5$ is not entirely surprising in view of the 
symmetries of the equations and the use of a 6d notation. 
What about the $S^5$? Could this be described by the $\Phi$ coordinates? 
In our description the  $SU(4)$ R-symmetry of ${\cal N}=4$ has been broken to (complexified) $SU(3)$. This suggest that one 
should not expect to find an $S^5$ because it is not consistent with the symmetries.  But it is interesting to note that
$\Phi$ can be identified with the (complexified) space of symmetric $SU(3)$ matrices, which is isomorphic to $SU(3)/SO(3)$ and is indeed 5 dimensional.

\section{Comments on the application to Yang-Mills theory}

In this note we studied the twistor construction of classical ${\cal N}=3$ super
Yang-Mills theory on the quadric submanifold $Q_{5|6}$ of $\cp^{3|3} \times
\cp^{3|3}$. We gave a reformulation of the twistor equations in six dimension,
and described the (anti)self-dual alpha- and beta-planes in a manifest superconformal invariant
notation. 
The superconformal symmetry naturally allows a modification of the twistor equations leading to 
an interesting connection with $AdS_5$ and its supersymmetric extension.
An important question is what this implies for the Yang-Mills theory, and 
whether our construction can be applied to the AdS/CFT correspondence. In this concluding section we will present some comments
regarding this questions for the purely bosonic twistor equations. A more complete investigation 
is left for future work.

Our six dimensional view on the twistor equations can be used to extend a 4d  gauge field  to
a 6d gauge field as follows. First one uses the fact that any 4d gauge field can be represented as a 
vector bundle over $Q_5$. Then by applying a generalization of Ward's construction to our 6d bosonic twistor equations
one obtains a Yang-Mills field in six dimensions. 
Indeed, according to Siegel, \cite{siegel} a 4d Yang-Mills field, not obeying any equations,
can be mapped on to a 6d gauge field satisfying 
\be
\label{siegel1}
X^AF_{ABC}=0
\ee
with
\be
F_{ABC}=X_{[A}F_{BC]}
\ee
where $A=1,\ldots 6$ are the 6d space-time indices. Here we followed the notation of \cite{siegel}.
This equation expresses the integrability of the gauge field along
alpha- and beta-planes. Thus a vector bundle on $Q_5$ is related through a generalized twistor construction to a 
solution of (\ref{siegel1}). The Yang-Mills field equations are in this notation
\be
\label{siegel2}
\nabla^A F_{ABC}=0
\ee 
To get a solution equation (\ref{siegel2}) one has again to locally extend the 
vector bundle to $\cp^3\times\cp^3$. 
An interesting observation in this context is 
that the Anti-de-Sitter space that we found can be related to the coordinates $(y,z)\in\bm_4\times\bm_4$ 
introduced in \cite{wittenold} and described in section 2. There we wrote the Minkowski coordinate
as $x={1\over 2}(y+z)$.
By applying the same line of thought to our AdS description we write
$$
X_{\ab}={1\over 2}(Y_{\a\b}+Z_{\a\b})\qquad\mbox{with}\qquad Y_{\a\b} V^{\b}=0,\qquad
U_\a  Z^{\a\b}=0.
$$
with $Y^2=Z^2=0$. Hence up to rescaling we have $Y=(1,y^2,y)$ and
$Z=(1,z^2,z)$ with $(y,z)\in \bm_4\times\bm_4$. 
The modified twistor equations for $X_{\a\b}$ imply
\be
Y^{\a\b }U_{\b}=2\zeta V^{\a},\qquad\qquad  Z_{\a\b} V^{\b}=2\zeta U_{\a}.
\ee
These equations are consistent provided that $Y^{\a\b} Z_{\a\b}=-8\zeta^2$. In 
terms of $z$ and $y$ this gives $(z-y)^2=4\zeta^2$. In other words, the
parameter $\zeta$ can be  interpreted as the distance $w^2=(z-y)^2$ between the two
points $y$ and $z$. This observation suggests that the gauge field on the AdS submanifold can be obtained from the
connections $\nabla_y$ and $\nabla_z$ in a point $(y,z)\in \bm_4\times\bm_4$.
However, a slightly confusing point is the following. In section 2 it was noted that an extension 
away from the diagonal in $\bm_4\times\bm_4$ requires the Yang-Mills equations to be satisfied. But here we just argued
that we can extend any gauge field to 6 dimensions. We believe the resolution is that for 
the construction of the 6d gauge field it is not necessary that $\nabla_y$ and $\nabla_z$ mutually commute.

We end with some final comments.
All these equations have presumably a supersymmetric extension.
In that case one does also obtain the super Yang-Mills equations: integrability
along the superalpha- and beta-planes gives the familiar constraints of $N=3$ supersymmetry which imply the equations 
of motion.  It would be interesting to work this out in detail in our formalism. We leave this for future work. 
Our work may be helpful in making a connection to the topological 
B-model on the quadric, and represent the Yang-Mills theory as a holomorphic Chern-Simons theory on the quadric. 
It has been suggested that such a formulation requires the introduction of harmonic superspace \cite{nair}.

Finally, it would be interesting to examine if the ${\cal N}=4$ gauge 
theory amplitudes can be formulated in terms of the twistor space $Q_{5|6}$. 
Although the amplitudes are in the weak coupling region, it is possible that 
one can find a sign of the quadric by performing a kind of Fourier 
transformation. Such a formulation would have the advantage of being 
symmetric in both helicities. 

\subsection*{Acknowledgment}
We would like to thank Jan de Boer for helpful discussions. This work is
supported in part by the Stichting FOM.

\end{document}